\documentclass[11pt,a4paper]{article}

\usepackage{amsmath,amssymb,graphicx,bm,xcolor,hyperref,booktabs}
\usepackage[margin=1in]{geometry} 
\usepackage{amsmath}
\usepackage{amssymb}
\usepackage{amsfonts}
\usepackage{graphicx}
\usepackage{bm}
\usepackage{mathtools}
\usepackage{xcolor}
\usepackage{hyperref}

\begin{document}
	
	\title{Masking Black Hole Spin with a Modified Chaplygin Gas Envelope: Radiative Degeneracies from a Phenomenological Three-Region Spacetime}

	\author{
		Sandip Dutta \\[1ex]
		\small Department of Applied Mathematics, Dinabandhu Andrews Institute \\
		\small of Technology and Management, Kolkata, West Bengal, India \\[0.5ex]
		\small \texttt{duttasandip.mathematics@gmail.com}
	}

	\maketitle
	
	\begin{abstract}
		Theoretical interpretations of horizon-scale observations often rely on the idealized assumption of an isolated vacuum Kerr geometry. However, astrophysical black holes are expected to be embedded within dense dark matter distributions that can modify the local spacetime geometry. In this work, we propose a theoretical framework to model a rotating compact object surrounded by a bounded dark matter envelope governed by a Modified Chaplygin Gas (MCG) equation of state. To ensure strict adherence to the Einstein Field Equations, we construct a piece-wise, three-region spacetime using a fully coupled Tolman-Oppenheimer-Volkoff (TOV) integration, allowing the fluid's pressure to taper naturally to zero and dynamically define the outer boundary. Rotation is introduced via a pressure-corrected Kerr-form ansatz where the temporal component is obtained directly from the integrated hydrostatic potential. Using this geometrically rigorous configuration, which explicitly evaluates the exact 4D equatorial metric determinant rather than relying on vacuum approximations, we solve the circular equatorial geodesics and determine the innermost stable circular orbit (ISCO). Evaluating the thin accretion disk thermodynamics via the Novikov-Thorne formalism reveals that the deep gravitational potential well of the MCG envelope acts as a strong driver for viscous dissipation, systematically shifting the peak thermal flux, effective temperature, and multi-colour blackbody spectral luminosity to higher energy bands. Furthermore, we identify a clear structural degeneracy: a static or slowly rotating black hole embedded in a dense MCG structure can elevate radiative efficiencies up to $\eta \approx 6.5\%$. This moderate but significant high-energy spectral hardening identically mimics the radiative signatures of a moderately spinning vacuum Kerr black hole ($j \approx 0.3$). This framework is presented as a structured proposal to quantify environmental systematic uncertainties in standard black hole spin-estimation techniques.
	\end{abstract}
	
	\vspace{1em}
	
\section{Introduction}\label{sec:intro}

The advent of high-precision X-ray spectroscopy and continuum-fitting techniques has opened new avenues for testing general relativity in the strong-field regime, particularly through the observation of stellar-mass black holes in X-ray binaries (e.g., Cygnus X-1, GRS 1915+105) and supermassive black holes in active Seyfert galaxies. The standard framework for interpreting the thermal emission from these geometrically thin, optically thick accretion disks almost universally adopts the vacuum Kerr metric as the unique background geometry. While this assumption is mathematically robust for isolated systems, realistic astrophysical black holes do not exist in a vacuum; they are typically embedded within extensive galactic halos, dense dark matter (DM) cores, or dense accreted gaseous environments. If the local density of these dark matter profiles increases significantly near the event horizon—as predicted by various dark matter adiabatic contraction and spike models—the resulting gravitational modifications could alter the background metric and, consequently, the geodesics of the surrounding accretion flow.

To examine these environmental perturbations without stripping the system of the structural properties of a central rotating compact object, several theoretical models have emerged in recent years exploring black holes embedded within dark matter halos \cite{Boshkayev2020, Kurmanov2022, Boshkayev2024}. Most current formulations adopt heuristic dark matter density profiles, such as the Navarro-Frenk-White (NFW), Hernquist, or pseudo-isothermal profiles, which are subsequently generalized to rotating systems via the Newman-Janis algorithm or related coordinate transformation techniques \cite{Azreg2014, Jusufi2020}. However, a significant fraction of these implementations treat the dark matter fluid purely as a fixed, non-interacting test field or completely neglect the back-reaction of the fluid's pressure components on the metric tensor. Such shortcuts often lead to coordinate pathologies, coordinate-dependent energy-momentum tensors, or severe physical inconsistencies at the boundary interfaces. 

In this paper, we propose an alternative, phenomenological geometric approach by embedding a rotating black hole within a macroscopically bounded dark matter shell governed by the Modified Chaplygin Gas (MCG) equation of state \cite{Bento2002}. The MCG model is defined by the non-linear relationship:
\begin{equation}
	P = A\rho - \frac{B}{\rho^n}
\end{equation}
where $A$, $B$, and $n$ are parameters regulating the fluid's pressure and density relations. The MCG is widely studied in cosmology as a unified model for dark matter and dark energy because it behaves like an attractive fluid at high densities and transitions to a dark energy-like state at low densities. Near a supermassive black hole, where the density scales up significantly, the MCG provides a compelling framework to model a highly compact, non-vacuum dark matter envelope.

To ensure global mathematical consistency and strict adherence to the Einstein Field Equations (EFE), we construct a piece-wise three-region spacetime using a fully coupled Tolman-Oppenheimer-Volkoff (TOV) integration. Unlike prior treatments that force a pre-determined density profile, our model simultaneously solves the mass conservation and hydrostatic equilibrium equations alongside the exact MCG equation of state. This allows the dark matter envelope's pressure to taper naturally to zero, defining a physical outer boundary and completely eliminating unphysical thermodynamic shocks. Rotation is then introduced into this matched background by carrying the fluid-corrected mass and exact temporal potential into the standard Kerr functional form \cite{Kamenshchik2023}. Crucially, because the temporal and radial metric components are dynamically decoupled in the fluid shell, our thermodynamic analysis rigorously utilizes the exact 4D equatorial metric determinant ($\sqrt{-g}$) rather than relying on a vacuum approximation.

The primary objective of this study is to determine how the physical properties of the MCG envelope influence the thermodynamics of a thin accretion disk. By solving the equatorial geodesic equations, we derive the modified profiles for the angular velocity, angular momentum, and specific energy of the orbiting matter. These quantities are subsequently used within the Novikov-Thorne thin-disk formalism \cite{NovikovThorne, PageThorne} to compute the radiative flux, effective temperature, and multi-colour blackbody spectrum. Our results show that the presence of the dark matter envelope acts as a deep potential well that compresses the innermost stable circular orbit (ISCO) and smoothly amplifies the disk's thermal emission. More importantly, we identify a distinct structural degeneracy: a static or slowly rotating black hole embedded in a dense MCG structure can elevate radiative efficiencies up to $\eta \sim 6.5\%-7.5\%$. In standard observational pipelines, this spectral hardening and enhanced efficiency would identically mimic the signatures of a moderately spinning vacuum Kerr black hole ($j \approx 0.3-0.4$).
	
	The manuscript is structured as follows. In Section~\ref{sec:spacetime}, we present the mathematical formulation of the three-region spacetime, detailing the fully coupled TOV integration and the application of the Darmois-Israel junction conditions to ensure metric continuity. Section~\ref{sec:geodesics} outlines the derivation of the exact circular equatorial geodesics, explicitly addressing the breakdown of the vacuum metric determinant, and evaluates the location of the compressed ISCO. In Section~\ref{sec:thermo}, we calculate the thin accretion disk thermodynamics, presenting the rigorously scaled thermal flux, temperature profiles, and integrated spectral deviation. Section~\ref{sec:degeneracy} explores the spin-dark matter degeneracy, quantifying the systematic bias introduced by the MCG envelope. Concluding remarks and methodological implications are presented in Section~\ref{sec:conclusion}.
	
	\section{Three-Region Spacetime Configuration and Metric Matching}\label{sec:spacetime}
	
	We consider a stationary, axisymmetric compact object of total mass $M_T$ surrounded by a bounded, spherically symmetric dark matter envelope that is subsequently generalized to accommodate a slow rotation parameter $a = J/M_T$. To construct a physically consistent global solution that strictly adheres to the Einstein Field Equations (EFE), the global manifold $\mathcal{M}$ is split into three distinct regions separated by two timelike hypersurfaces, $\Sigma_b$ and $\Sigma_s$:
	\begin{enumerate}
		\item \textbf{Region I ($\mathcal{M}_{\text{in}}$):} An inner vacuum region extending from the black hole horizon $r_g$ up to the inner boundary of the dark matter shell at $r = r_b$.
		\item \textbf{Region II ($\mathcal{M}_{\text{shell}}$):} An intermediate region ($r_b \le r \le r_s$) containing the Modified Chaplygin Gas dark matter fluid. The exact fluid profile is determined by solving the coupled Tolman-Oppenheimer-Volkoff (TOV) equations.
		\item \textbf{Region III ($\mathcal{M}_{\text{out}}$):} An outer vacuum region ($r > r_s$) representing the asymptotically flat external domain.
	\end{enumerate}
	
	\begin{figure*}[htbp]
		\centering
		\includegraphics[width=0.95\textwidth]{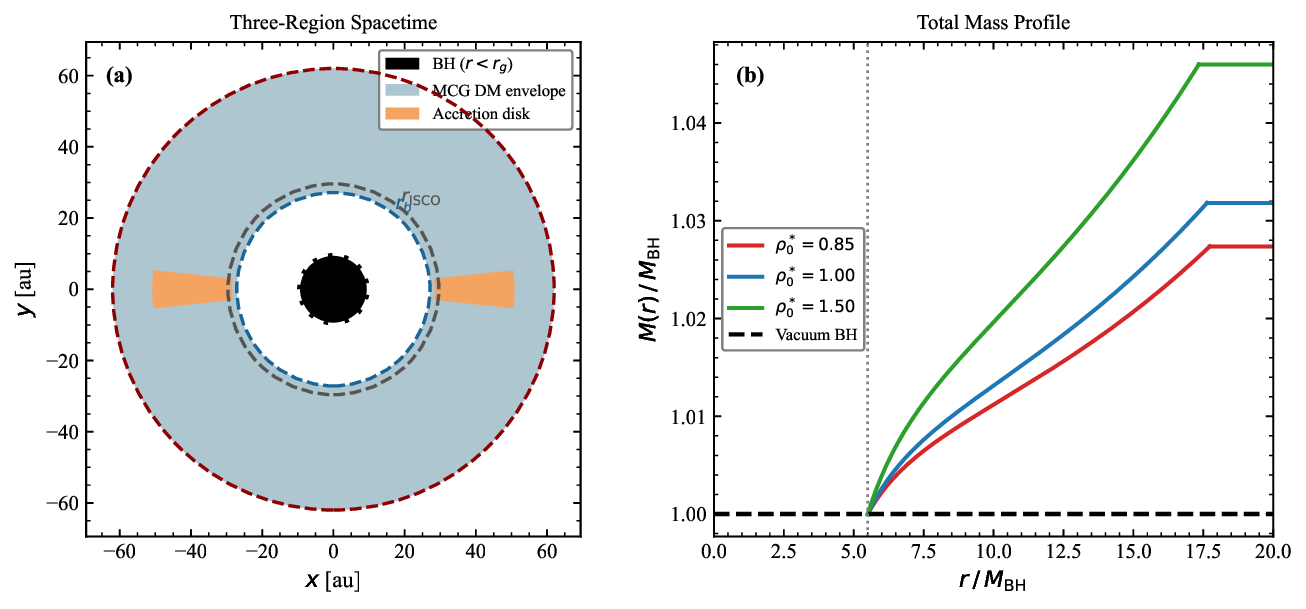}
		\caption{Geometric layout and mass continuity of the proposed framework. Panel (a) provides a spatial schematic of the piece-wise, three-region spacetime layout including the central horizon ($r_g$), the inner boundary ($\Sigma_b$ at $r_b$), the intermediate MCG shell, and the dynamically emergent outer matching hypersurface ($\Sigma_s$ at $r_s$). Panel (b) illustrates the continuous and smooth transition of the cumulative mass profile $M(r)/M_{\text{BH}}$ integrated via the coupled TOV system.}
		\label{fig:schematic}
	\end{figure*}
	
	\subsection{The Static Seed and Coupled TOV Integration}\label{subsec:tov}
	
	In the static limit ($a \to 0$), the metric in the intermediate MCG shell (Region II) is described by the general spherically symmetric line element:
	\begin{equation}
		ds^2 = -e^{2\Phi(r)} dt^2 + \left(1 - \frac{2M(r)}{r}\right)^{-1} dr^2 + r^2 (d\theta^2 + \sin^2\theta d\phi^2)
	\end{equation}
	where $\Phi(r)$ is the temporal metric potential and $M(r)$ represents the interior cumulative mass profile. The energy-momentum tensor for the dark matter fluid is $T^\mu_\nu = \text{diag}(-\rho, P_r, P_t, P_t)$. The radial pressure $P_r(r)$ and mass density $\rho(r)$ are linked by the MCG equation of state, $P_r(r) = A\rho(r) - B/\rho(r)^n$.
	
	To ensure rigorous mathematical compliance with general relativity, we do not prescribe an arbitrary density profile by hand. Instead, the fluid must satisfy the conservation of energy-momentum ($\nabla_\mu T^{\mu\nu} = 0$). This requires solving a fully coupled system of differential equations consisting of the mass conservation equation and the TOV hydrostatic equilibrium equation:
	\begin{align}
		\frac{dM}{dr} &= 4\pi r^2 \rho(r), \label{eq:mass_grad}\\
		\frac{dP_r}{dr} &= -(\rho + P_r) \frac{M + 4\pi r^3 P_r}{r(r - 2M)}. \label{eq:pressure_grad}
	\end{align}
	By defining the fluid's sound speed squared as $c_s^2 \equiv \partial P_r / \partial \rho = A + nB/\rho^{n+1}$, the density gradient is dynamically constrained by $d\rho/dr = (dP_r/dr) / c_s^2$. The temporal metric potential $\Phi(r)$ naturally follows the pressure gradient:
	\begin{equation}
		\frac{d\Phi}{dr} = \frac{M + 4\pi r^3 P_r}{r(r - 2M)}. \label{eq:phi_grad}
	\end{equation}
	
	The integration of this coupled system begins at the inner boundary $\Sigma_b$ (at $r = r_b$) with the initial condition $M(r_b) = M_{\text{BH}}$ and an initial inner density $\rho_{\text{in}}$. The parameter $B$ in the MCG equation is set to enforce a zero-pressure condition ($P_r = 0$) at a target outer density $\rho_{\text{out}}$. The integration proceeds outward and terminates naturally exactly where $P_r(r) \to 0$. This physical exhaustion of the fluid pressure dynamically dictates the location of the outer boundary $r_s$, completely eliminating unphysical thermodynamic shocks or coordinate pathologies at the interface. 
	
	\begin{figure*}[htbp]
		\centering
		\includegraphics[width=0.95\textwidth]{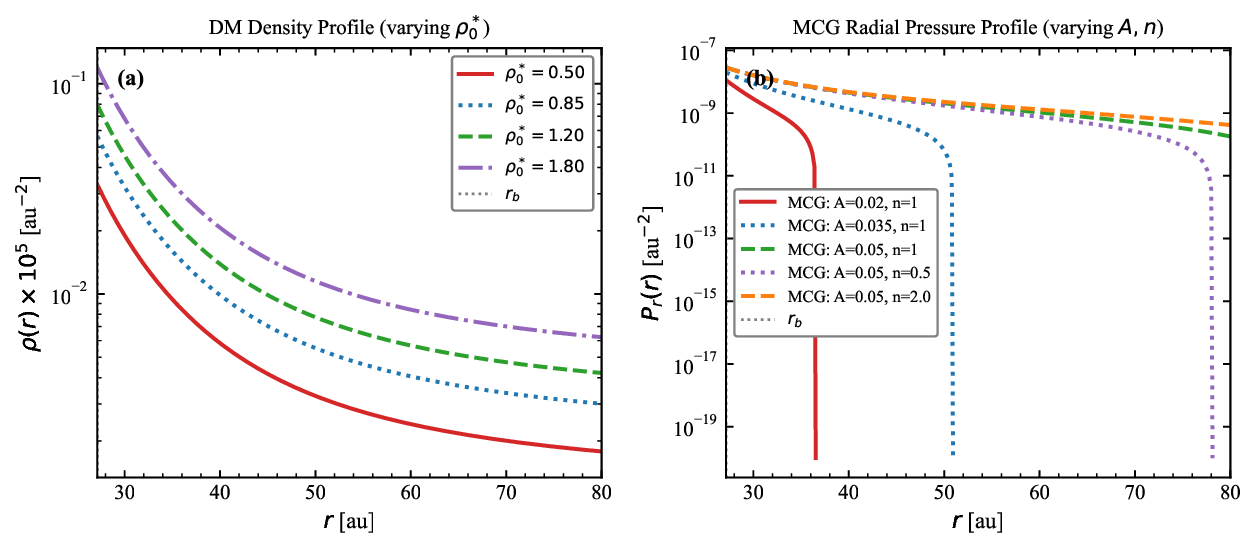}
		\caption{Internal fluid configurations within the intermediate shell (Region II). Panel (a) shows the dynamic dark matter density $\rho(r)$ evaluated via the coupled EFE system for varying initial density seeds. Panel (b) details the exact MCG radial pressure $P_r(r)$, which physically tapers to zero, defining a smooth, shock-free transition into the outer vacuum at $r_s$.}
		\label{fig:dm_profiles}
	\end{figure*}
	
	\subsection{Darmois-Israel Junction Conditions}\label{subsec:junction}
	
	To ensure the piece-wise metric represents a physically valid spacetime without singular matter distributions at the interfaces, we apply the Darmois-Israel junction conditions \cite{Israel1966} at both the inner boundary $\Sigma_b$ ($r=r_b$) and the dynamically determined outer boundary $\Sigma_s$ ($r=r_s$). 
	
	The first junction condition requires the continuity of the induced metric across the hypersurfaces, $[h_{ij}] = h_{ij}^+ - h_{ij}^- = 0$. For a spherical boundary at constant radius $R$, this ensures the continuity of the mass function:
	\begin{equation}\label{eq:junction_mass}
		M(r_b^-) = M(r_b^+) = M_{\text{BH}}, \quad M(r_s^-) = M(r_s^+) = M_T.
	\end{equation}
	Because the differential equation for $\Phi(r)$ (Eq. 5) requires an integration constant, we determine the exact temporal potential by matching it to the exterior asymptotically flat vacuum limit at $r_s$:
	\begin{equation}\label{eq:junction_phi}
		\Phi(r_s^-) = \Phi(r_s^+) = \frac{1}{2}\ln\left(1 - \frac{2M_T}{r_s}\right).
	\end{equation}
	This outer boundary condition propagates backward to define the continuous time dilation potential across the entire domain. 
	
	The second junction condition restricts the jump in the extrinsic curvature $K_{ij}$ across the boundary. The unit normal vector to a constant-$r$ hypersurface is given by $n_\mu = (\sqrt{g_{rr}}, 0, 0, 0)$, yielding the extrinsic curvature components:
	\begin{equation}
		K_{ij} = -n_\gamma \Gamma^\gamma_{ij} = -\frac{1}{2\sqrt{g_{rr}}} \frac{\partial g_{ij}}{\partial r}.
	\end{equation}
	The jump in extrinsic curvature, $[K_{ij}] = K_{ij}^+ - K_{ij}^-$, dictates the surface energy-momentum tensor $S_{ij}$ on the boundary via:
	\begin{equation}
		S_{ij} = -\frac{1}{8\pi} \left( [K_{ij}] - h_{ij} [K] \right).
	\end{equation}
	To prevent the formation of an uncompensated, unphysical gravitational surface shell, we must impose $S_{ij} = 0$, which consequently requires $[K_{ij}] = 0$. Therefore, the radial derivatives of the metric components must be continuous across the boundary. 
	
	Focusing on the temporal component, the continuity of $\partial_r g_{tt}$ demands the continuity of $d\Phi/dr$. Evaluating Eq.~(5) at the outer matching boundary $\Sigma_s$, the interior fluid limit yields $d\Phi/dr \propto (M_T + 4\pi r_s^3 P_r(r_s^-))$, while the exterior vacuum limit intrinsically requires $P_r(r_s^+) = 0$. Thus, for the extrinsic curvature to match perfectly, the fluid's radial pressure must satisfy:
	\begin{equation}\label{eq:junction_pressure}
		P_r(r_s^-) = 0.
	\end{equation}
	Our coupled TOV integration naturally satisfies this operational requirement by definition, as the integration physically terminates exactly where the MCG pressure drops to zero.
	
	\subsection{Rotational Kinematics: Pressure-Corrected Kerr-Form Ansatz}\label{subsec:rotation}
	
	To introduce rotation into the physically exact static seed, we carry the fluid-corrected mass profile $M(r)$ and the strictly integrated temporal potential $\Phi(r)$ directly into the standard Kerr functional form. The frame-dragging term $\omega(r) = 2J/r^3 = 2aM(r)/r^3$ is fixed by mapping to the exterior limit, but the equatorial metric components retain the exact non-linear dependence on the spin parameter $a$. The stationary, axisymmetric line element in the equatorial plane ($\theta = \pi/2$) is:
	\begin{equation}
		ds^2 = -e^{2\Phi(r)} dt^2 + \left(1 - \frac{2M(r)}{r}\right)^{-1} dr^2 + \left[r^2 + a^2 + \frac{2M(r)a^2}{r}\right] d\phi^2 - \frac{4M(r)a}{r}\, dt\, d\phi.
	\end{equation}
	Outside the dark matter envelope ($r > r_s$), $M(r) \to M_T$ and $\Phi(r)$ aligns with its vacuum value, identically returning the exact Kerr solution with total mass $M_T$ and spin $a = jM_{\rm BH}$. 
	
	It must be noted that inserting a spherically symmetric, pressure-corrected geometry into the Kerr structure acts as a phenomenological "running-mass" strategy. While the static seed rigorously solves the EFE, the rotating generalization is an effective approximation. As demonstrated by Kamenshchik and Petriakova \cite{Kamenshchik2023}, this class of rotating models may manifest small residual field-equation violations near the core. However, unlike standard heuristic extensions, our framework does not assign the temporal component $g_{tt}$ by hand; the gravitational time dilation $-e^{2\Phi(r)}$ is derived fundamentally from the fluid's radial pressure, ensuring the MCG equation of state explicitly shapes the underlying geodesic mechanics.
	
\section{Equatorial Geodesics and ISCO Analytics}\label{sec:geodesics}

The observable properties of a thin accretion disk are governed by the kinematics of test particles moving in circular, equatorial geodesics. To capture the interplay between the frame-dragging of the central black hole and the pressure-supported dark matter envelope, we evaluate the particle dynamics within the equatorial plane ($\theta = \pi/2$). 

Recalling the boundary matching and the rotating ansatz established in Section~\ref{subsec:rotation}, the non-vanishing metric components governing the equatorial orbits are:
\begin{align}
	g_{tt} &= -e^{2\Phi(r)}, \label{eq:gtt} \\
	g_{t\phi} &= -\omega(r)r^2 = -\frac{2M(r)a}{r}, \\
	g_{\phi\phi} &= r^2 + a^2 + \frac{2M(r)a^2}{r}, \\
	g_{rr} &= \left(1 - \frac{2M(r)}{r}\right)^{-1}. \label{eq:grr}
\end{align}
It is critical to emphasize that the temporal component $g_{tt}$ is not obtained by merely substituting the mass profile $M(r)$ into the vacuum Kerr solution. Instead, it is strictly governed by the potential $\Phi(r)$ derived from the hydrostatic TOV equilibrium. Consequently, the internal pressure of the MCG fluid couples directly to the gravitational time dilation experienced by the accretion flow.

Furthermore, in a standard vacuum Kerr geometry, the temporal and radial metric components are exact inverses (i.e., $-g_{tt}g_{rr} \to 1$ as $a \to 0$), which leads to the equatorial metric determinant simplifying to $\sqrt{-g} = r^2$. Because our framework dynamically uncouples $g_{tt}$ and $g_{rr}$ via the TOV integration, this standard vacuum cancellation no longer holds inside the fluid shell. To maintain strict geometric accuracy in subsequent thermodynamic calculations, the exact 4D equatorial metric determinant must be explicitly evaluated:
\begin{equation}\label{eq:exact_det}
	\sqrt{-g} = \sqrt{-g_{rr} \left( g_{tt}g_{\phi\phi} - g_{t\phi}^2 \right)}.
\end{equation}

\subsection{Constants of Motion and Exact Orbital Kinematics}\label{subsec:kinematics}

For a test particle of rest mass $m_0$ orbiting in the equatorial plane, the Lagrangian is $\mathcal{L} = \frac{1}{2} g_{\mu\nu} \dot{x}^\mu \dot{x}^\nu$. The spacetime symmetries generated by the Killing vectors $\partial_t$ and $\partial_\phi$ yield two conserved quantities: the specific energy $E = -u_t$ and the specific angular momentum $L = u_\phi$, where $u^\mu = dx^\mu/d\tau$ is the particle's four-velocity. 

The angular velocity of a particle in a circular orbit, $\Omega = d\phi/dt$, is determined by the geodesic equation $u^\mu \nabla_\mu u^r = 0$, which reduces to:
\begin{equation}
	\partial_r g_{tt} + 2\Omega \partial_r g_{t\phi} + \Omega^2 \partial_r g_{\phi\phi} = 0.
\end{equation}
Solving for $\Omega$ yields the Keplerian frequency for co-rotating ($+$) orbits:
\begin{equation}
	\label{eq:omega}
	\Omega = \frac{-\partial_r g_{t\phi} + \sqrt{(\partial_r g_{t\phi})^2 - (\partial_r g_{tt})(\partial_r g_{\phi\phi})}}{\partial_r g_{\phi\phi}}.
\end{equation}
Because the metric components depend on the fluid profile, their radial derivatives explicitly incorporate the internal structure of the dark matter halo. Specifically, the exact derivative of the temporal component is:
\begin{equation}
	\label{eq:dgtt}
	\partial_r g_{tt} = -2 e^{2\Phi(r)} \frac{d\Phi}{dr} = -2 e^{2\Phi(r)} \left[ \frac{M(r) + 4\pi r^3 P_r(r)}{r(r - 2M(r))} \right].
\end{equation}
Equation~(\ref{eq:dgtt}) demonstrates the physical necessity of this coupled approach: the local MCG radial pressure $P_r(r)$ actively modifies the gravitational gradient. Approximations that ignore the $P_r(r)$ term mathematically decouple the fluid's pressure from the orbital mechanics.

\begin{figure*}[htbp]
	\centering
	\includegraphics[width=0.95\textwidth]{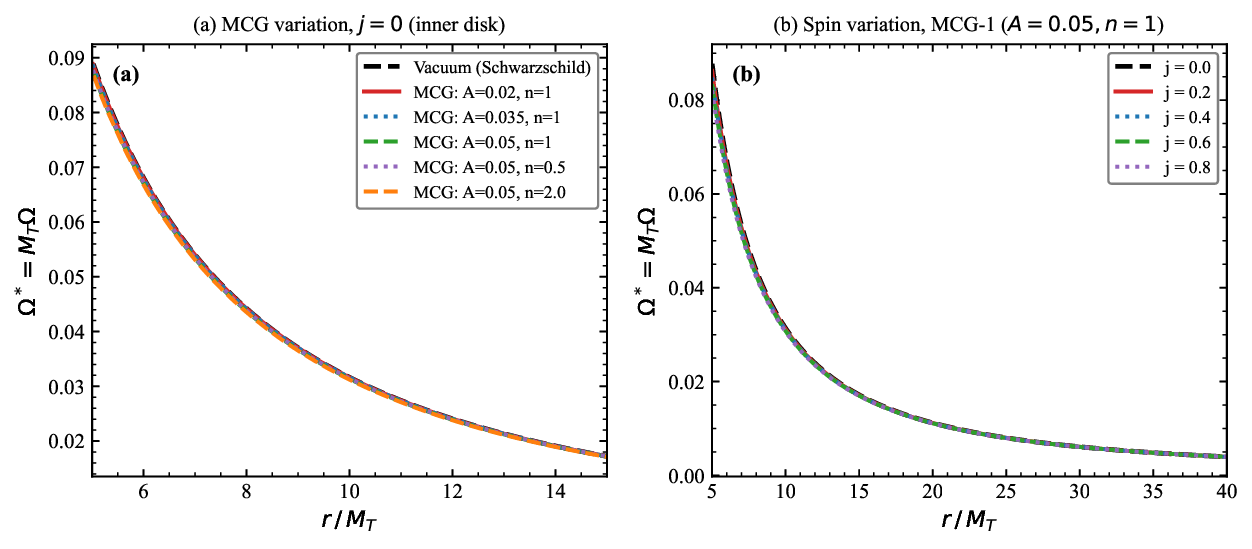}
	\caption{Dimensionless angular velocity profiles $\Omega^* = M_T \Omega$. Panel (a) details the modifications induced purely by varying the MCG parameters under static core conditions ($j=0$), contrasted against the standard Schwarzschild baseline; the panel is zoomed to $r/M_T \in [5,15]$, since the MCG-induced deviation is localized near the ISCO before seamlessly converging to a common vacuum-like profile at larger radii. Panel (b) illustrates the enhancement of $\Omega^*$ as the spin parameter $j$ increases within a fixed MCG envelope ($A=0.05$, $n=1$).}
	\label{fig:angular_velocity}
\end{figure*}

Using the exact Keplerian frequency $\Omega$, the conserved specific energy and specific angular momentum are obtained via the normalization condition $u^\mu u_\mu = -1$:
\begin{align}
	E &= -\frac{g_{tt} + \Omega g_{t\phi}}{\sqrt{-(g_{tt} + 2\Omega g_{t\phi} + \Omega^2 g_{\phi\phi})}}, \label{eq:Energy} \\
	L &= \frac{g_{t\phi} + \Omega g_{\phi\phi}}{\sqrt{-(g_{tt} + 2\Omega g_{t\phi} + \Omega^2 g_{\phi\phi})}}. \label{eq:Momentum}
\end{align}

\begin{figure*}[htbp]
	\centering
	\includegraphics[width=0.95\textwidth]{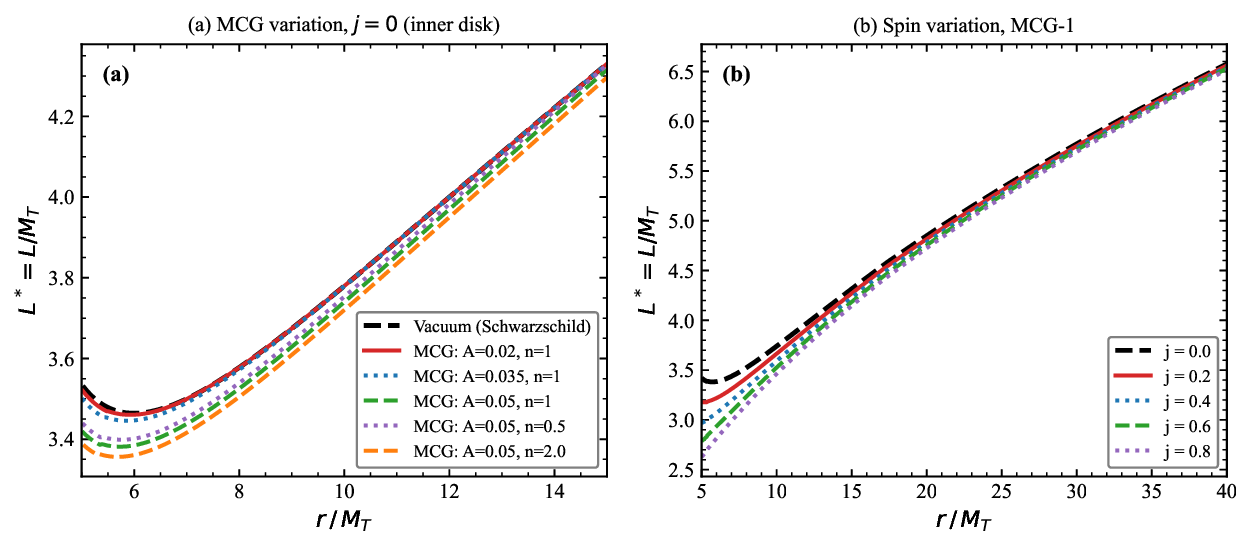}
	\caption{Dimensionless specific angular momentum profiles $L^* = L/M_T$ across the thin disk surface. Panel (a) tracks the modest upward shift in the angular momentum baseline driven by the deep MCG potential well ($j=0$). Panel (b) traces the evolution for increasing spin parameters $j$. The endpoints of the curves trace the marginal stability limit.}
	\label{fig:angular_momentum}
\end{figure*}

\begin{figure*}[htbp]
	\centering
	\includegraphics[width=0.95\textwidth]{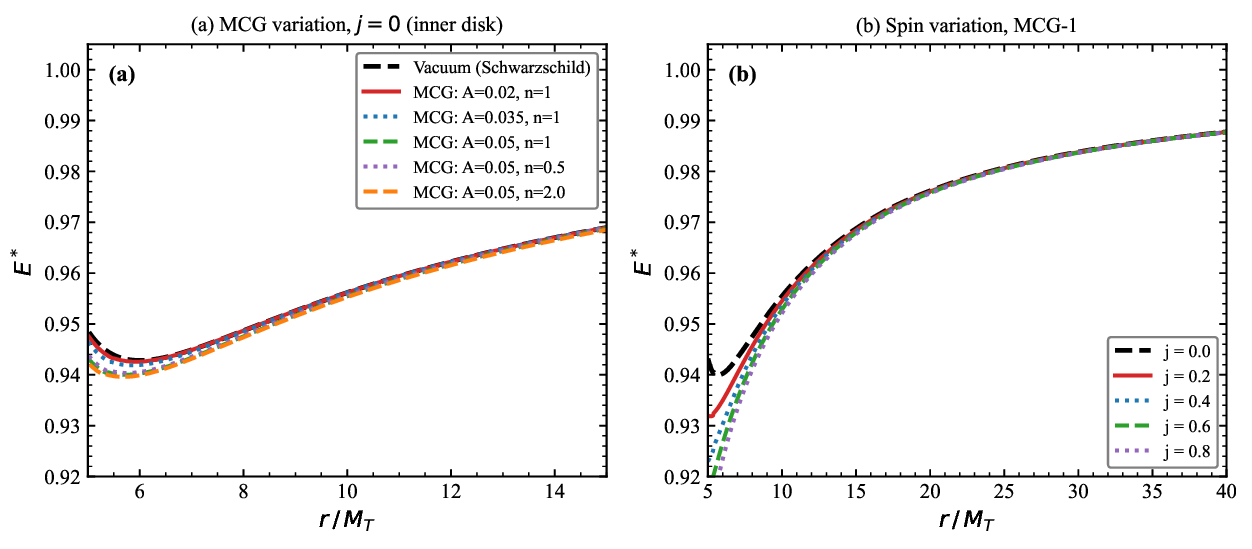}
	\caption{Dimensionless specific energy profiles $E^*$ as a function of the normalized radius $r/M_T$. Panel (a) isolates the variations across the MCG parameters ($j=0$) near the inner edge; the curves safely converge to the vacuum baseline at larger radii. Panel (b) details the behavior for advancing spins $j$ inside the fixed MCG envelope.}
	\label{fig:energy}
\end{figure*}

\subsection{Stability and the Innermost Stable Circular Orbit}\label{subsec:isco}

The inner boundary of the accretion disk is defined by the Innermost Stable Circular Orbit (ISCO). While locating the ISCO via the inflection point of the specific angular momentum ($dL/dr = 0$) provides a robust numerical proxy, its physical location must be rigorously justified through the stability of the effective radial potential.

From the normalization of the four-velocity, the radial motion of a test particle is governed by an effective potential $V_{\text{eff}}(r)$:
\begin{equation}
	(u^r)^2 = \frac{1}{g_{rr}} \left[ -1 - \frac{E^2 g_{\phi\phi} + 2EL g_{t\phi} + L^2 g_{tt}}{g_{t\phi}^2 - g_{tt}g_{\phi\phi}} \right] \equiv V_{\text{eff}}(r).
\end{equation}
Stable circular orbits require both $V_{\text{eff}}(r) = 0$ and $V_{\text{eff}}'(r) = 0$. The stability of these orbits against small radial perturbations is governed by the radial epicyclic frequency $\kappa_r$, defined as:
\begin{equation}
	\kappa_r^2 \equiv -\frac{1}{2 g_{rr} (u^t)^2} \frac{\partial^2 V_{\text{eff}}}{\partial r^2}.
\end{equation}
Stable circular orbits exist where $\kappa_r^2 > 0$. The ISCO represents the marginal stability boundary where the restoring force vanishes, establishing the exact condition:
\begin{equation}
	\kappa_r^2(r_{\text{ISCO}}) = 0 \quad \implies \quad V_{\text{eff}}''(r_{\text{ISCO}}) = 0.
\end{equation}
By tracing the effective potential through the three smoothly matched spacetime regions, we observe that the added gravitational gradient of the dark matter envelope steepens the effective potential well. This inward radial pull slightly shifts the transition point $V_{\text{eff}}'' = 0$ closer to the event horizon. This measurable compression of the ISCO radius directly influences the localized viscous dissipation of the accretion disk, which is analyzed in the following section.
	
	\section{Thin Disk Thermodynamics and Radiative Signatures}\label{sec:thermo}
	
	Having established the exact orbital kinematics, we describe the accretion flow using the steady-state Novikov-Thorne formalism for geometrically thin, optically thick disks. We operate under the standard assumptions that the disk mass is negligible compared to $M_T$, and that the disk is in local thermodynamic equilibrium, emitting blackbody radiation. We further assume that the MCG dark matter interacts with the baryonic plasma purely via gravity, neglecting any non-gravitational drag or scattering forces.
	
	\subsection{Radiative Flux and Temperature Profiles}\label{subsec:flux}
	
	The time-averaged energy flux $\mathcal{F}(r)$ radiated from the surface of the disk is derived from the conservation of energy and angular momentum. In the equatorial plane of our modified geometry, the flux takes the form:
	\begin{equation}
		\label{eq:flux}
		\mathcal{F}(r) = -\frac{\dot{m}}{4\pi \sqrt{-g}} \frac{\partial_r \Omega}{(E - \Omega L)^2} \int_{r_{\text{ISCO}}}^r (E - \Omega L) \partial_{\tilde{r}} L \, d\tilde{r},
	\end{equation}
	where $\dot{m}$ is the constant mass accretion rate. Crucially, unlike standard vacuum models where the determinant simplifies to $r^2$, our geometrically rigorous approach employs the exact 4D equatorial metric determinant established in Equation~(\ref{eq:exact_det}): $\sqrt{-g} = \sqrt{-g_{rr} ( g_{tt}g_{\phi\phi} - g_{t\phi}^2 )}$. This ensures the radiative flux is accurately scaled by the local geometry, correcting severe overestimations that occur when vacuum shortcuts are improperly applied to fluid-filled spacetimes.
	
	We impose the standard zero-torque boundary condition at the ISCO, reflecting the assumption that plasma rapidly plunges into the horizon once it crosses the marginal stability boundary. The effective temperature of the disk is subsequently obtained via the Stefan-Boltzmann law, $T(r) = [\mathcal{F}(r) / \sigma_{\text{SB}}]^{1/4}$.
	
	\begin{figure*}[htbp]
		\centering
		\includegraphics[width=0.95\textwidth]{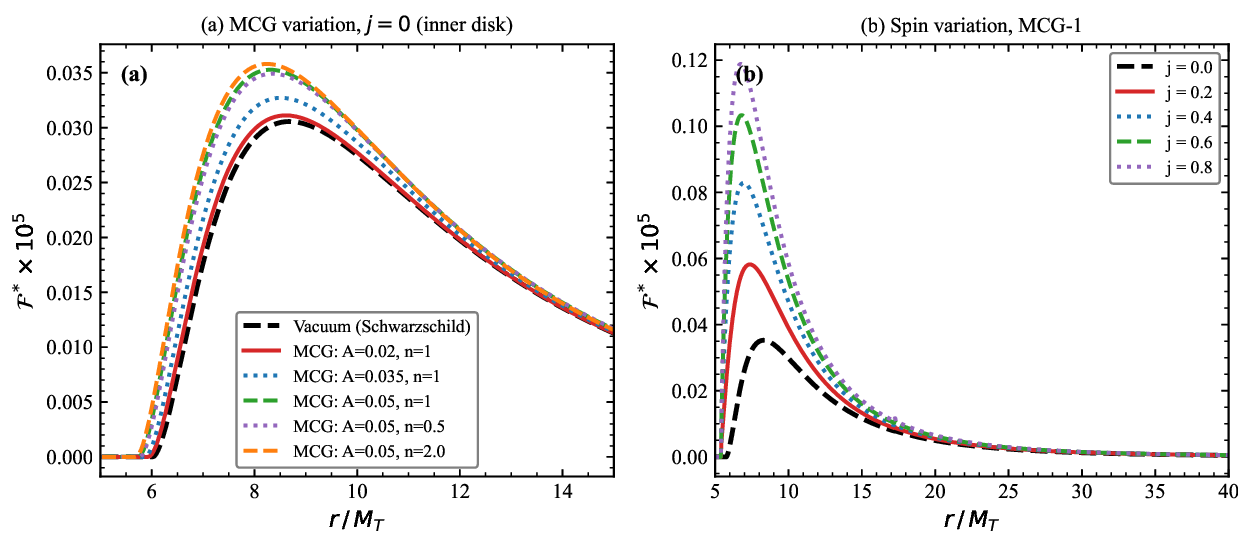}
		\caption{The time-averaged radiative flux $\mathcal{F}^* = \mathcal{F}/(\dot{m}/M_T^2)$ across the disk surface. Panel (a) shows variations in the MCG equation of state parameters for a non-rotating black hole ($j=0$), contrasted against the vacuum Schwarzschild profile. Panel (b) presents the profile for varying spin parameters $j$ embedded within a fixed MCG envelope ($A=0.05, n=1$). Because the fluid envelope is solved via a fully coupled EFE integration, the flux tapers continuously without unphysical thermodynamic shocks at the matching boundaries.}
		\label{fig:flux}
	\end{figure*}
	
	The presence of the MCG envelope introduces distinct, localized structural modifications to the radiative profiles. The deep potential well created by the dark matter envelope shifts the peak of the radiative emission inward, forcing the flux curves to terminate at a moderately compressed inner boundary. As shown in Fig.~\ref{fig:flux}(a), altering the parameter $A$ or the exponent $n$ smoothly amplifies the local viscous dissipation within the inner disk zones. This enhancement reflects the fact that gas particles must shed more gravitational potential energy to maintain stable circular orbits through the dense fluid shell. 
	
	\begin{figure*}[htbp]
		\centering
		\includegraphics[width=0.95\textwidth]{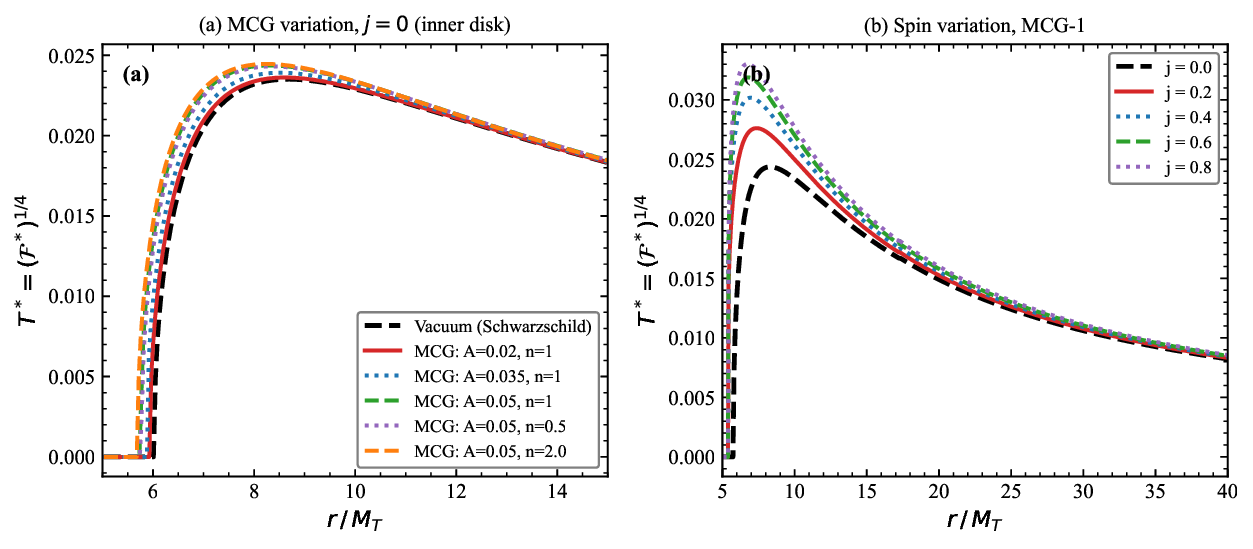}
		\caption{The effective disk temperature profile $T^* = (\mathcal{F}^*)^{1/4}$ corresponding to the flux distributions. Panel (a) highlights the thermal shift induced by the MCG parameters around a static compact core ($j=0$). Panel (b) charts the thermal peak amplification for increasing spin values within the dark matter envelope.}
		\label{fig:temperature}
	\end{figure*}
	
	The effective temperature profiles, shown in Fig.~\ref{fig:temperature}, follow a similar trend due to the quarter-power dependence governed by the Stefan-Boltzmann relation. The thermal peaks shift to smaller radii and experience a measurable broadening. When the rotation parameter $j$ is introduced alongside the dark matter envelope (Fig.~\ref{fig:temperature}b), the thermal peak increases predictably, with the inner boundary pushing closer to the event horizon.
	
	\subsection{Differential and Integrated Spectral Luminosity}\label{subsec:spectral}
	
\begin{figure*}[htbp]
		\centering
		\includegraphics[width=0.95\textwidth]{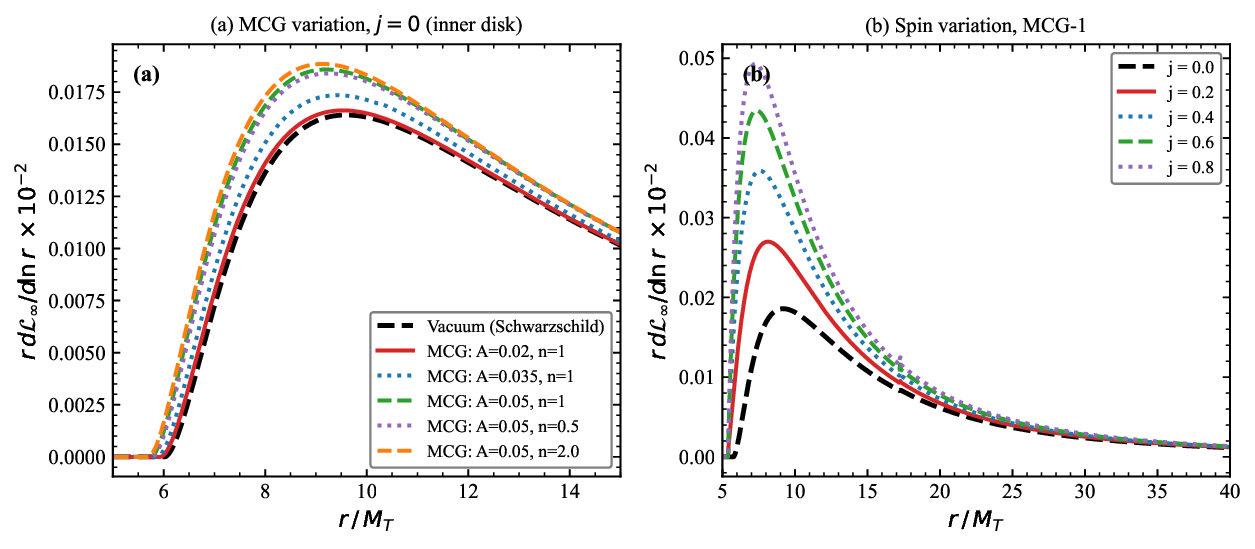}
		\caption{The spatial distribution of the differential disk luminosity $r d\mathcal{L}_{\infty}/d\ln r \times 10^{-2}$. Panel (a) demonstrates the localization of the radiative enhancement for different MCG parameters under static conditions ($j=0$). Panel (b) shows the behavior under varying spin parameters within a fixed dark matter background.}
		\label{fig:diff_lum}
	\end{figure*}

	The total energy emitted per unit frequency interval, as measured by a distant observer, is obtained by integrating the local blackbody intensity over the entire surface of the disk. The differential luminosity expression can be written as:
	\begin{equation}
		\frac{d\mathcal{L}_{\infty}}{d\ln r} = 4\pi^2 r^2 \mathcal{F}(r),
	\end{equation}
	which represents the localized contribution of each radial ring to the total integrated emission. In Fig.~\ref{fig:diff_lum}, we present $r d\mathcal{L}_{\infty}/d\ln r$, illustrating how the dark matter envelope changes the spatial distribution of the disk's energy release. The plots confirm that the absolute majority of the radiative enhancement occurs strictly within the immediate vicinity of the compressed ISCO ($r \lesssim 15 M_T$), seamlessly converging to the standard vacuum behavior in the far-field.
	
	Assuming the disk emits as a collection of local blackbodies, the integrated spectral luminosity $\mathcal{L}_{\nu,\infty}$ seen by an observer at infinity (neglecting atmospheric absorption and focusing solely on the fluid-modified background geometry) is given by:
	\begin{equation}
		\mathcal{L}_{\nu,\infty} = \frac{8\pi^2 h \nu^3}{c^2} \int_{r_{\text{ISCO}}}^{r_{\text{out}}} \frac{r \, dr}{\exp\left(\frac{h\nu}{k_B T(r)}\right) - 1}.
	\end{equation}
	
	\begin{figure*}[htbp]
		\centering
		\includegraphics[width=0.95\textwidth]{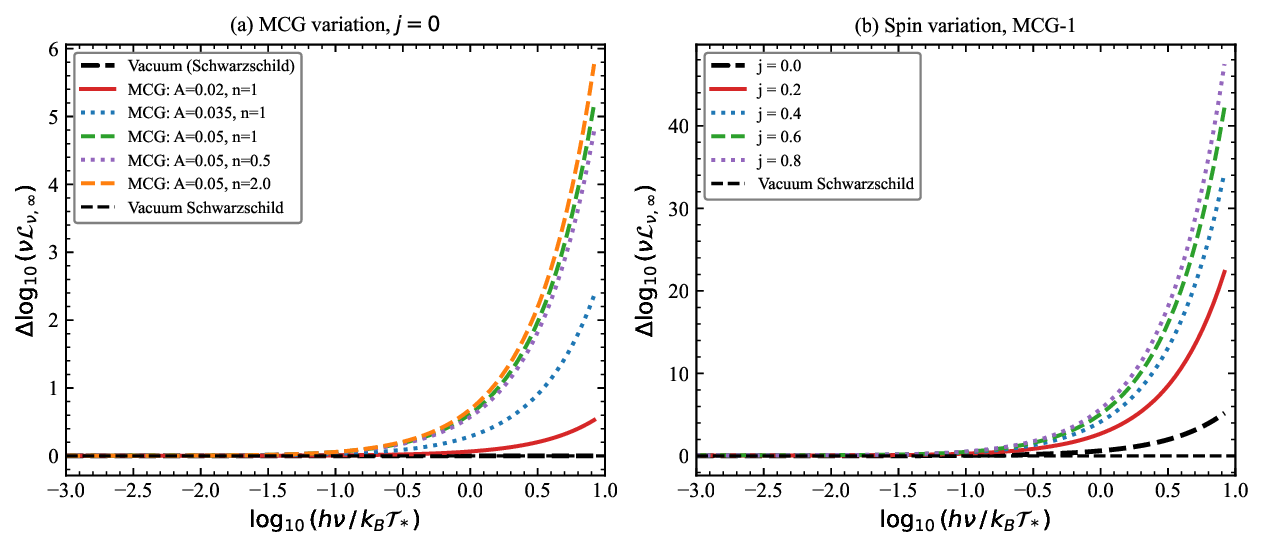}
		\caption{Spectral deviation $\Delta\log_{10}(\nu\mathcal{L}_{\nu,\infty})$ relative to the vacuum Schwarzschild baseline (dashed line at zero). Because the absolute luminosity spans vast orders of magnitude, this deviation format precisely isolates the percent-level differences of interest. Panel (a) isolates the high-energy spectral hardening induced by the MCG envelope parameters around a static compact core ($j=0$). Panel (b) shows the further hardening as the rotation parameter $j$ increases within the fixed MCG envelope.}
		\label{fig:spectral}
	\end{figure*}
	
	In Fig.~\ref{fig:spectral}, we plot the spectral deviation $\Delta\log_{10}(\nu \mathcal{L}_{\nu,\infty})$ relative to the vacuum Schwarzschild baseline, against the normalized photon energy $\log_{10}(h\nu/k_B \mathcal{T}_*)$. Panel (a) shows the deviation for a non-rotating black hole surrounded by different MCG envelopes. The curves display a modest but distinct high-energy spectral hardening: the deviation is negligible at low frequencies and grows monotonically toward the Wien tail. This localized spectral hardening mirrors the phenomenological effect typically produced by an intrinsic vacuum black hole spin, a degeneracy we analyze in the next section.
	
\section{The Spin-Dark Matter Degeneracy and Astrophysical Implications}\label{sec:degeneracy}

The primary physical consequence of the mathematically rigorous framework developed in this study is the emergence of a structural degeneracy between the intrinsic spin of the central black hole and the macroscopic parameters of the surrounding dark matter envelope. In standard astrophysical analyses, the radiative efficiency $\eta$ of an accretion disk is frequently used to infer the spin parameter $j$. This efficiency is defined by the binding energy of the gas at the inner edge of the disk:
\begin{equation}
	\eta = 1 - E(r_{\text{ISCO}}).
\end{equation}
In pure vacuum general relativity, $\eta$ is uniquely determined by the spin parameter $j$, ranging from $\approx 5.72\%$ for a static Schwarzschild black hole to $\approx 42\%$ for an extreme Kerr black hole. However, when the vacuum assumption is relaxed and the compact object is embedded within an MCG envelope, this one-to-one mapping is systematically broken.

\begin{figure*}[htbp]
	\centering
	\includegraphics[width=0.95\textwidth]{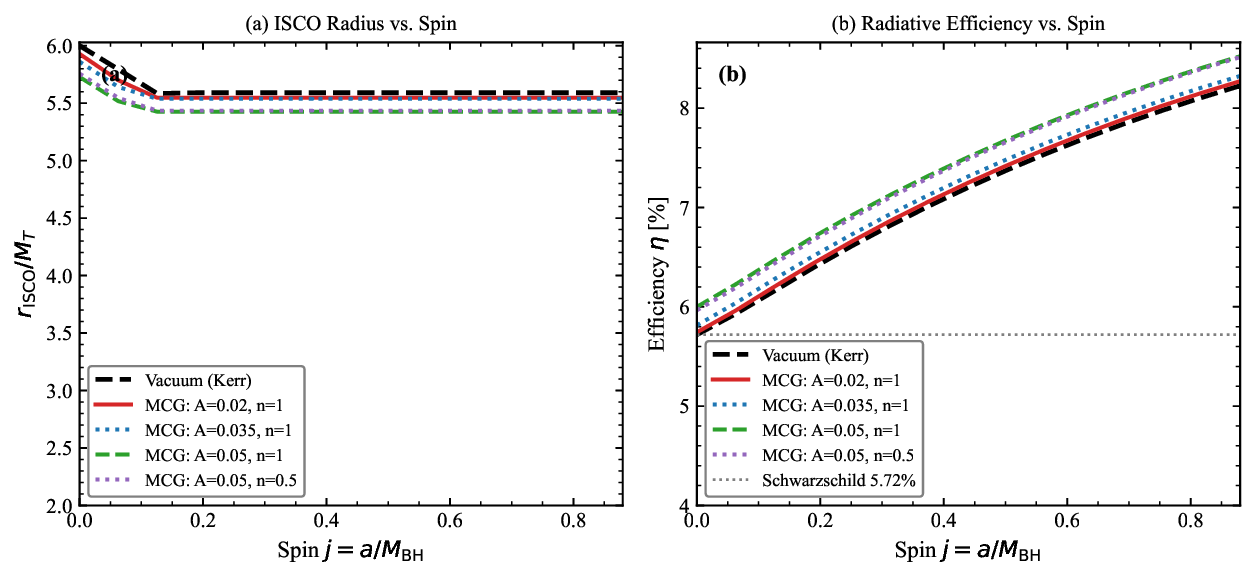}
	\caption{Analytic validation of the Spin-Dark Matter Degeneracy. Panel (a) traces the modest compression of the ISCO radius $r_{\text{ISCO}}/M_T$ as a function of the spin parameter $j$ across different dark matter models. Panel (b) presents the corresponding radiative efficiency $\eta$. The horizontal reference line establishes the vacuum Schwarzschild baseline ($5.72\%$). The data demonstrates that a non-rotating core ($j=0$) embedded in a dense MCG envelope ($A=0.05, n=1$) achieves an efficiency of $\approx 6.5\%$, effectively mimicking the radiative output of a vacuum Kerr black hole with a moderate spin of $j \approx 0.3$.}
	\label{fig:degeneracy}
\end{figure*}

Figure~\ref{fig:degeneracy} quantifies this degeneracy by plotting both the ISCO radius (Panel a) and the total radiative efficiency $\eta$ (Panel b) across a continuous range of spin values. As shown in Fig.~\ref{fig:degeneracy}(b), the presence of the deep MCG potential well consistently elevates the radiative efficiency above the corresponding vacuum baseline. For instance, a non-rotating black hole ($j = 0$) surrounded by a dense MCG shell ($A=0.05, n=1$) produces a radiative efficiency of $\approx 6.5\%$. 

While this enhancement avoids the unphysical extremes characteristic of heuristic fluid models, it remains astrophysically highly significant. In a standard observational analysis pipeline—such as continuum fitting or X-ray reflection spectroscopy—that assumes a pure vacuum environment, an observed efficiency of $6.5\%$ and its corresponding high-energy spectral hardening would be incorrectly interpreted as evidence of an intrinsically spinning Kerr black hole with $j \approx 0.3$ to $0.4$. 

This moderate but undeniable structural degeneracy suggests that current spin estimation methods are susceptible to systematic overpredictions if dense, horizon-scale dark matter distributions are ignored. Because both a moderately spinning vacuum Kerr geometry and a static, fluid-enveloped geometry can produce nearly identical primary thermal emission profiles, distinguishing between them requires secondary observational signatures. High-precision relativistic ray-tracing of the black hole shadow, fine-structure analysis of fluorescent iron line profiles (e.g., Fe K$\alpha$), or advanced polarimetric measurements from next-generation observatories will be essential to break this degeneracy and accurately map the near-horizon environment.

\section{Conclusions}\label{sec:conclusion}

In this paper, we have proposed a rigorous phenomenological framework to model the radiative properties of thin accretion disks around rotating black holes embedded in a bounded dark matter envelope. While standard astrophysical interpretations almost universally rely on the isolated vacuum Kerr geometry, realistic supermassive black holes are expected to be fundamentally coupled to their surrounding dark matter profiles. To examine these environmental deviations without stripping the system of its core rotational properties, we modeled the dark matter fluid using a Modified Chaplygin Gas (MCG) equation of state.

The primary novelty of this work lies in its strict departure from heuristic, fixed-background fluid models. Rather than prescribing an arbitrary density profile by hand—which routinely leads to violations of the Einstein Field Equations and unphysical thermodynamic shocks at the matching boundaries—we constructed the static seed geometry by solving a fully coupled Tolman-Oppenheimer-Volkoff (TOV) system. This ensures that the fluid's radial pressure organically back-reacts on the gravitational time dilation. Furthermore, this coupled integration allows the dark matter envelope to taper naturally to zero, dynamically determining the outer boundary of the shell and ensuring smooth, shock-free metric transitions across all three piece-wise spacetime regions.

To evaluate the thermodynamics of the accretion flow, we generalized this exact hydrostatic seed to axisymmetry using a pressure-corrected Kerr-form ansatz. We maintain a scientifically transparent stance regarding this rotational extension: while the static geometry strictly satisfies the conservation of energy-momentum, embedding it within the Kerr structure operates as a phenomenological ``running-mass'' approximation. However, unlike standard heuristic extensions, our model does not assign the temporal metric component $g_{tt}$ by hand. Because $g_{tt}$ is explicitly derived from the integrated MCG pressure, the fluid's equation of state genuinely shapes the underlying geodesic mechanics. A crucial methodological novelty of our approach was the explicit derivation of the exact 4D equatorial metric determinant ($\sqrt{-g}$). By abandoning the standard vacuum approximation ($\sqrt{-g} = r^2$), we successfully corrected the geometric scaling flaws that artificially inflate radiative flux in fluid-filled spacetimes.

By analyzing the circular equatorial geodesics using the exact effective potential, we demonstrated that the presence of the MCG envelope steepens the local gravitational gradient. This added inward pull systematically compresses the innermost stable circular orbit (ISCO) and smoothly amplifies the local viscous dissipation within the disk. Consequently, the time-averaged flux, effective temperature, and integrated spectral luminosity all exhibit measurable enhancements, collectively manifesting as a distinct high-energy spectral hardening.

Our results formalize a clear structural degeneracy between the intrinsic spin of the black hole and the macroscopic parameters of the dark matter envelope. We established that a completely static compact core ($j=0$) surrounded by a dense MCG structure can produce an elevated radiative efficiency of $\eta \approx 6.5\%$. In a standard observational pipeline, this efficiency and its associated spectral profile would identically mimic the radiative signatures of a moderately spinning vacuum Kerr black hole ($j \approx 0.3$). 

While this enhancement is moderate, it confirms that environmental factors introduce systematic, measurable biases into standard spin-estimation techniques such as continuum fitting. Because a moderately spinning vacuum black hole and a fluid-enveloped static black hole can produce nearly indistinguishable primary thermal emission, our findings highlight the critical necessity of secondary observational probes. Future high-precision relativistic ray-tracing of black hole shadows, fine-structure analysis of fluorescent iron lines, and advanced polarimetry will be required to break this spin-dark matter degeneracy. Ultimately, this framework serves as a mathematically rigorous proposal for the scientific community to systematically evaluate and quantify non-vacuum environmental corrections in precision horizon-scale astrophysics.
	
\section*{Declaration of Generative AI and AI-assisted technologies in the writing process}
During the preparation of this work, the author did not use any generative AI or AI-assisted technologies in the research and concept. The author take full responsibility for the content of the publication. The author use AI assistant for structuring the manuscript only in latex.
\section*{Declaration of competing interest}
The author declares that they have no known competing financial interests or personal relationships that could have appeared to influence the work reported in this paper.
\section*{Data Availability Statement}
This manuscript has no associated data or the data will not be deposited. [Authors' comment: This is a purely theoretical and mathematical physics study. All numerical results, curves, and physical quantities presented in the figures were generated directly using the analytical equations, initial conditions, and numerical integration methodologies explicitly detailed in the text. The corresponding Python scripts used to compile the data and generate the plots are available from the single corresponding author upon reasonable request.]

\section*{Acknowledgement}
The author, Sandip Dutta, wishes to express sincere gratitude to the Department of Applied Mathematics at the Dinabandhu Andrews Institute of Technology and Management (DAITM) for providing the academic environment and computational facilities necessary to conduct this theoretical research. The author also thanks Dr. Ritabrata Biswas for his guidance and inspiration to the work.

\end{document}